\documentstyle[12pt,subeqnarray,epsf,repko]{article}
\def\hsp{\mbox{\rule{14pt}{0pt}}}
\begin{document}
\def\om{\omega}
\def\Om{\Omega}
\def\mn{\mu\nu}
\def\al{\alpha}
\def\BM#1{{\boldmath
\mathchoice{\hbox{$\displaystyle#1$}}
            {\hbox{$\textstyle#1$}}
            {\hbox{$\scriptstyle#1$}}
            {\hbox{$\scriptscriptstyle#1$}}}}
\begin{flushright}
\hfill{\small UTEXAS-HEP-00-03}\\
\hfill{\small MSUHEP-00330}
\end{flushright}
\begin{center}

{ {\large\bf Neutrino-photon Scattering in a Magnetic Field}}

\vspace{10pt}

Duane A. Dicus$^1$ and Wayne W. Repko$^2$

\vspace{10pt}

{\small\it $^1$Center for Particle Physics, The University of Texas at Austin, Austin,
Texas 78712} \\

\vspace{10pt} {\small\it $^2$Department of Physics and Astronomy, Michigan State
University, East Lansing, MI 48824}

\end{center}

\vspace{24pt}

\begin{abstract}

Previous calculations of photon-neutrino scattering in a constant background
magnetic field are extended and corrected.
\end{abstract}

\baselineskip=18pt
\parskip=8pt
\section{\large\bf Introduction}

For massless neutrinos, the leading contribution to low energy neutrino-photon scattering
vanishes because the amplitude can be written in the form of a vector or axial-vector
current coupling to the photons. The largest low energy interaction between photons and
neutrinos is therefore $\gamma\nu\rightarrow\gamma\gamma\nu$ and the crossed channels.
Rates for these processes have been calculated, for energies less than the electron mass,
$m_e$, by using an effective action \cite{dad1} and, for larger energies, by a
straight-forward calculation of box diagrams \cite{dad2}.

In the presence of magnetic fields neutrino photon elastic scattering does exist because
the magnetic field gives an additional interaction -- it effectively replaces one of the
photons in $\gamma\nu\rightarrow\gamma\gamma\nu$ with a zero momentum insertion.  This
too has been calculated; for low energies using the effective action for the $2\to 3$
processes \cite{Sha} and, at higher energies, by calculating triangle diagrams in the
external field \cite{chyi,chyi1}.

Our purpose here is to criticize mildly and extend slightly these external
field results. In particular the effective action paper \cite{Sha} gave results
for only one channel  ($\gamma\gamma\rightarrow\nu\overline\nu$). We extend
this to all channels with $B$ perpendicular and parallel for each channel.
Results of the more general calculation of Ref.\,\cite{chyi} are given in three
figures. We believe the curves in these figures are somewhat in error; for
example, they sometimes disagree at low energies with our extended effective
action results.  Both references \cite{Sha} and \cite{chyi} express their
results in terms of the critical magnetic field $B_c=m_e^2/e = 4.41\times
10^{13}$ gauss.  In each case the authors have taken $e$ to be in
unrationalized units, $e^2=1/137$. whereas it should have been taken to be in
natural units $e^2=4\pi/137$.  Thus all their results are too large by a factor
of $4\pi$.

\section{\large\bf Neutrino-photon cross sections in a uniform magnetic field}

The calculation of Ref.\,\cite{Sha} uses the effective interaction given in
Ref.\,\cite{dad1}. In this interaction, as in the Euler-Heisenberg Lagrangian upon which
it is based, $\alpha$ equals $e^2/4\pi$ so the coupling of the magnetic field has been
taken as this $e$.  As discussed in Ref.\,\cite{Raf} the critical field, when written as
$m_e^2/e$, uses $e$ in these same natural units. The value of the critical field,
$4.41\times 10^{13}$ gauss, means that in natural (or rationalized) units 1 tesla
$=195\;{\rm eV}^2$. The result of Ref.\,\cite{Sha} agrees in every respect with the
result given below for $\gamma\gamma\rightarrow \nu\overline\nu$ with the external field
perpendicular to the direction of the photons except in the replacement of $B^2_c$ by
$m_e^4/\alpha$ where it should be replaced by $m_e^4/4\pi\alpha$.  Since the curve
$\gamma\gamma\rightarrow \nu\overline\nu$ in Fig. 2 of Ref.\,\cite{chyi} agrees, at low
energy, with the result of Ref.\,\cite{Sha} these authors too must have made the wrong
substitution for $B_c$.

\subsection{\normalsize\bf Low energy interactions}

The explicit calculations at low energy, using the effective action, are unremarkable and
give the following results.  For $\gamma\gamma\rightarrow\nu\overline\nu$ and
$\nu\overline\nu\rightarrow\gamma\gamma$ the cross sections with center of mass energy
$\omega$ and scattering angle $z$  are given by
\begin{equation}
\frac{d\sigma}{dz}=\sigma_0\;\frac{A}{4\pi}\;\left( \frac{\omega}{m_e}\right)^6\left(
\frac{B}{B_c}\right)^2\left[2312\left(1-N^2_1\right)+816\;N_1N_2z-N^2_2(744+72z^2)\right]\,,
\end{equation}
where $A=1$ for $\gamma\gamma\rightarrow\nu\overline\nu$ and 1/2 for $\nu\overline\nu
\rightarrow\gamma\gamma$.  For electron neutrinos, including both the $W$ and $Z$-boson
contributions,  $a=1/2 + 2 \sin^2\theta_W$  and
\begin{equation}
\sigma_0=\frac{G_F^2\alpha^2a^2 m^2_e}{(180)^2\pi^2} =2.12\times10^{-54} {\rm cm^2}\,,
\end{equation}
where $G_F$ is the Fermi constant and $\theta_W$ is the weak mixing angle. $N_1$ and
$N_2$ depend on whether the $B$ field is parallel or perpendicular to the direction of
the initial particles and are given in Table I.

The cross section for $\gamma\nu\rightarrow\gamma\nu$ is given by
\begin{equation}
\frac{d\sigma}{dz}=\frac{\sigma_0}{4\pi}\left( \frac{\omega}{m_e}\right)^6\left(
\frac{B}{B_c}\right)^2\left[1211-143z^2-\left(N^2_1 +N^2_2\right) (896+62z^2)
+N_1N_2z(859-11z^2)\right]\,
\end{equation}
where terms odd in $z$ have been dropped.  Using the $N_1$ and $N_2$ values and
completing the integral over $z$ gives
\begin{equation}\label{sigB}
\sigma=\frac{\sigma_0}{4\pi}\;\left(\frac{\omega}{m_e}\right)^6
\left(\frac{B}{B_c}\right)^2 {\BM X}\,,
\end{equation}
where the $\BM X$ values for the six cases are given in Table I.  Note that the result
for $\gamma\gamma\rightarrow\nu\overline\nu$ with perpendicular $B$ agrees with Ref.
\cite{Sha} if the $4\pi$ is ignored.  Also note that the differential cross sections for
parallel $B$ go as $1-z^2$ as they must.

The ${\BM X}$ values show that the cross sections for $\gamma\nu\rightarrow\gamma\nu$
differ by a factor of about 4 depending on whether the magnetic field is parallel or
perpendicular.  The cross sections for $\nu\overline\nu\rightarrow\gamma\gamma$ differ by
a factor of about 200.  These differences are not present in the figures of
Ref.\,\cite{chyi}.

\subsection{\normalsize\bf Interactions at arbitrary center of mass energies}

The authors of Refs.\,\cite{chyi, chyi1} do a nice job of deriving the lowest
order constant magnetic field contribution to photon-neutrino scattering, both
from the correction to the electron propagator and from the phase factor. Their
expression for the matrix element \cite{chyi1} seems entirely correct with the
exception of a relative minus sign in the tensor which multiplies the
coefficient function $C_{11}$. The authors of Ref.\,\cite{chyi1} inform us that
this is a typographical error \cite{lin}. We agree with all the coefficient
functions $C_1,\cdots,C_{11}$.

The results of our calculation of neutrino-photon scattering in a constant magnetic field
following the methods of Refs.\,\cite{chyi} and \cite{chyi1} are shown in Figures 1, 2,
and 3 for the three channels.  All calculations are for $B=0.1 B_c$ as is the case in
Ref.\,\cite{chyi}.  At low energies the cross sections agree almost exactly with
Eq.\,(\ref{sigB}). Note the relative factor of about 4 between the magnetic field
directions for $\gamma\nu\rightarrow\gamma\nu$ and the similar factor of about 200 for
$\nu\overline\nu \rightarrow\gamma\gamma$.  Also note the infrared divergence at
$\omega=m_e$ in the $\gamma\gamma\leftrightarrow\nu\overline\nu$ channels where the
magnetic field interaction acts as a zero energy insertion on an external leg.

For large energy all of the cross sections grow as $\omega^2$ except for
$\gamma\gamma\rightarrow\nu\overline\nu$ with $B$ parallel to the photon direction, which
decreases.  After summation over photon polarizations, the cross section for this channel
can be written as
\begin{eqnarray}
\sigma &=& A_1F^{\mu\nu}F_{\mu\nu}+A_2(k^\mu_1\;F_{\mu\alpha}F^{\alpha\nu}\;k_{1\nu}
+k_2^\mu\; F_{\mu\alpha}\;F^{\alpha\nu}\;k_{2\nu}) \nonumber\\ & &
+A_3\;k_1^\mu\;F_{\mu\alpha}\;F^{\alpha\nu}\;
k_{2\nu}+A_4(k_1^\mu\;F_{\mu\nu}\;k^\nu_2)^2\,,
\end{eqnarray}
where $F^{\mu\nu}$ is the field strength of the magnetic field and $k_1$ and $k_2$ are
the momenta of the photons.  The $A_i$ are combinations of the coefficient functions
$C_1,\,C_2,\,\dots\,,C_{11}$. For a magnetic field in a direction $\vec{n}$ this reduces to
\begin{equation}
\sigma \sim  2A_1 - \omega^2\;A_2\;[2-(\hat{k}_1\cdot \vec n)^2-(\hat{k}_2\cdot\vec
n)^2]+\omega^2A_3\;[1+ \widehat{k}_1\cdot\vec n\,\hat{k}_2\cdot\vec n\;]\,.
\end{equation}
Thus for $\gamma\gamma\rightarrow\nu\overline\nu$ with $\vec n$ parallel to $\hat{k}_1$
and $\hat{k}_2$ the coefficients of $A_2$ and $A_3$ vanish. $A_1$, which in this channel
depends only on $C_1$, falls as $\ln^2(\omega^2/m_e^2)/\omega^2$ for $\omega >> m_e$,
giving the decrease at large $\omega$ seen in Fig.\,(\ref{ggnnb}). $A_2$ and $A_3$ do not
vanish for $\vec n$ perpendicular to $\hat{k}_1$ and $\hat{k}_2$ nor do they vanish for
either $B$ direction in the other channels.

\section{\large\bf Conclusions}

\begin{figure}[h]
\hspace{1.5in} \epsfysize=2.7in \epsfbox{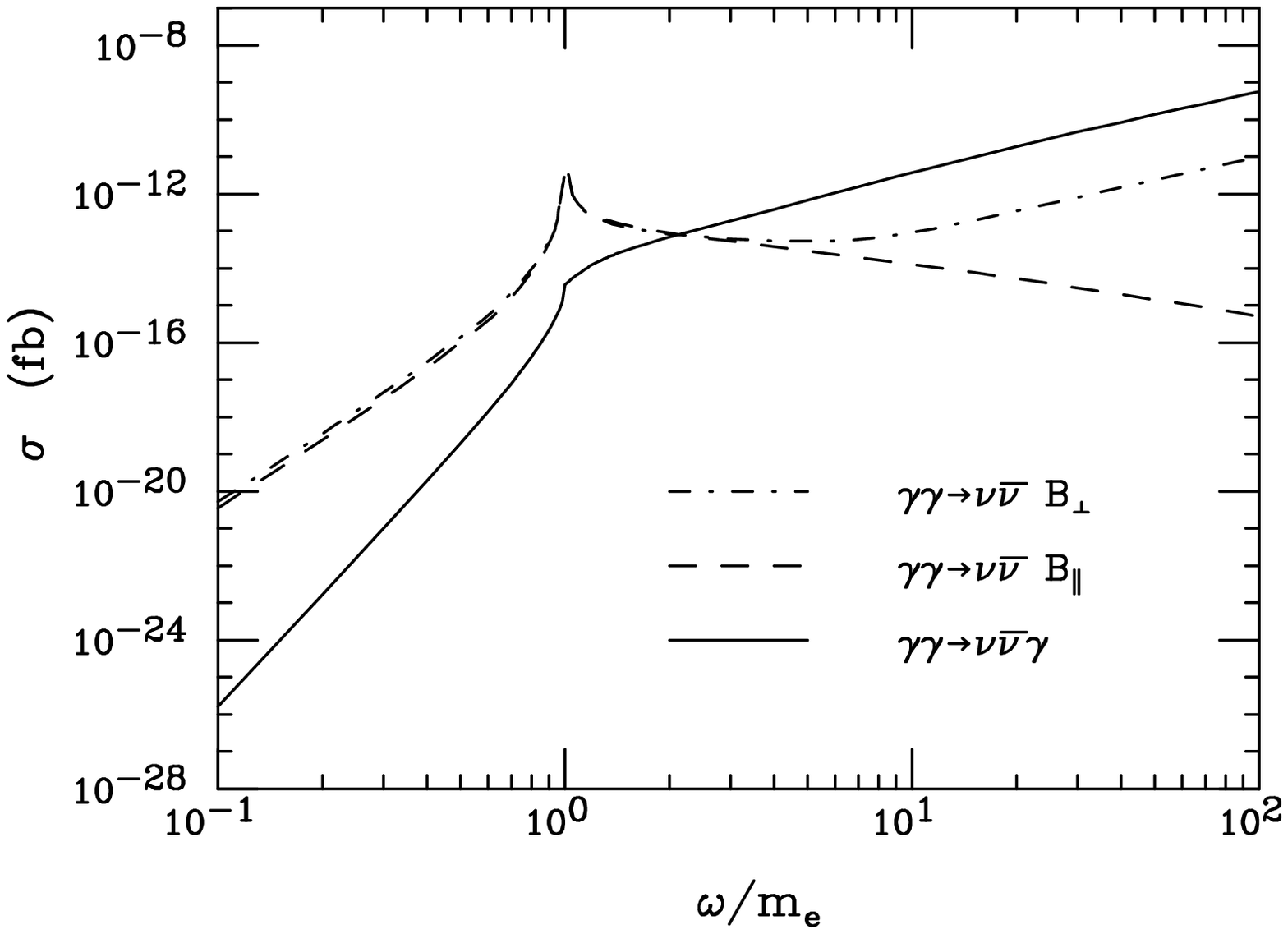}
\caption{\footnotesize The cross section $\sigma(\gamma\gamma\protect\to\nu\bar{\nu})$ in
the presence of a constant magnetic field $B$ is shown as a function of the photon energy
$\omega$. Two magnet field orientations, parallel and perpendicular to the direction
incoming photons in the center of mass are indicated. The solid line is the cross section
for the $2\protect\to 3$ process $\gamma\gamma\protect\to\nu\bar{\nu}\gamma$.\label{ggnnb}}
\end{figure}

\begin{figure}[h]
\hspace{1.5in} \epsfysize=2.7in \epsfbox{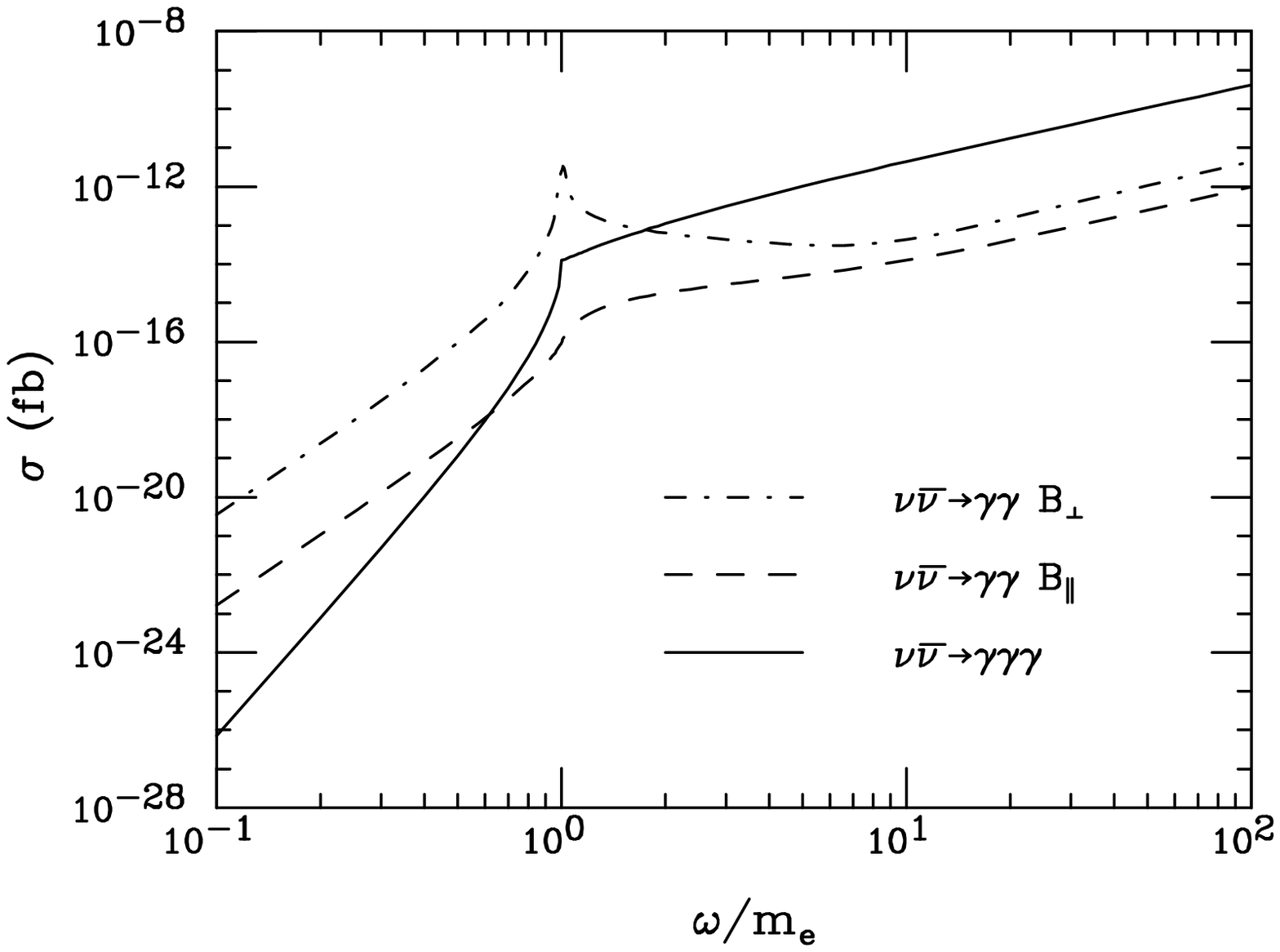}
\caption{\footnotesize The cross section $\sigma(\nu\bar{\nu}\protect\to\gamma\gamma)$ in
the presence of a constant magnetic field $B$ is shown as a function of the photon energy
$\omega$. Two magnet field orientations, parallel and perpendicular to the direction
incoming photons in the center of mass are indicated. The solid line is the cross section
for the $2\protect\to 3$ process $\nu\bar{\nu}\protect\to\gamma\gamma\gamma$. \label{nnggb}}
\end{figure}

\begin{figure}[h]
\hspace{1.5in}
\epsfysize=2.7in \epsfbox{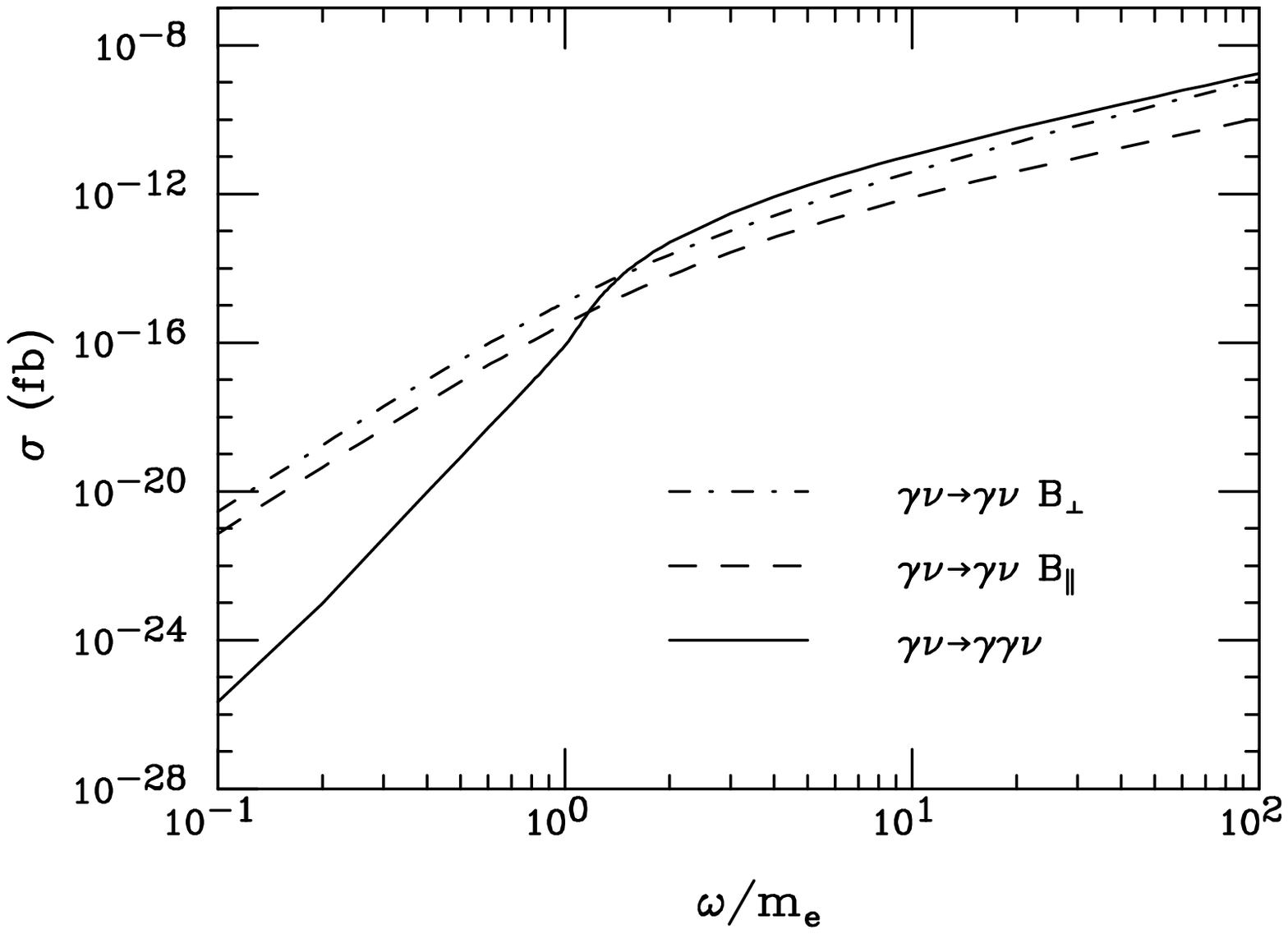}
\caption{\footnotesize The cross section $\sigma(\gamma\nu\protect\to\gamma\nu)$ in the
presence of a constant magnetic field $B$ is shown as a function of the photon energy $\omega$.
Two magnet field orientations, parallel and perpendicular to the direction incoming photons
in the center of mass are indicated. The solid line is the cross section for the
$2\protect\to 3$ process $\gamma\nu\protect\to\gamma\gamma\nu$. \label{gngnb}}
\end{figure}

  We have extended the low energy effective Lagrangian calculation of
Ref.\,\cite{Sha} to include all channels and directions of the magnetic field B. In
Figs.\,(\ref{ggnnb}-\ref{gngnb}), we have given correct values for the cross sections at
all energies, which, while shown for $B = 0.1B_c$, can be scaled by $B^2$.  (Of course,
following Ref.\,\cite{chyi} and \cite{chyi1}, we have essentially done perturbation
theory in $B/B_c$ so our results are not valid for much larger $B$.)  All calculations
were performed for electron neutrinos but this too can be easily changed by using $a =
1/2 - 2\sin^2\theta_W$ for the other neutrino types.

  It turns out that neutrino photon scattering in an external magnetic
field is less important, relative to the $2\to 3$ processes, than previously thought [4].
Since the effects of these processes on stars was already calculated [4] to be small we
have not repeated the determination of the stellar energy loss rates or mean free paths.

\begin{center}
\section*{\large\bf Acknowledgement}
\end{center}
 We would like thank Chung Kao, Vic Teplitz and Roberto Vega for helpful conversations
and R. Shaisultanov and G.-L. Lin for comments. This work was supported in part
by the National Science Foundation under grant PHY-9802439 and by the
Department of Energy under Grant No. DE-FG13-93ER40757.

\newpage

\begin{center}
\section*{\large\bf References}
\end{center}


\pagebreak

\begin{center}
\section*{\large\bf Tables}
\end{center}

\begin{table}[h]
\begin{center}
\begin{tabular}{|c|c|c|r|}\hline
Process & $N_1$ &  $N_2$ & ${\BM X}$\hsp \\ \hline
$(\gamma\gamma\rightarrow\nu\overline\nu)_\bot$& ~~~~$y/\sqrt{2}$~~~~
&0 &9248/3 \\ \hline
$(\gamma\gamma\rightarrow\nu\overline\nu)_{\|}$ & z & 1 &$ 6272/3 $\\ \hline
$(\nu\overline\nu\rightarrow\gamma\gamma)_\bot$ & 0 & $y/\sqrt{2}$ &
10296/5\\ \hline
$(\nu\overline\nu\rightarrow\gamma\gamma)_{\|}$& 1 & $z$ & 48/5\\ \hline
$(\gamma\nu\rightarrow\gamma\nu)_\bot$& 0 & ~~~~$y/\sqrt{2}~~~~$ & 25816/15\\ \hline
$(\gamma\nu\rightarrow\gamma\nu)_{\|}$& 1 & $z$ &  6592/15 \\ \hline
\end{tabular}
\caption{\label{tab1}
The subscript $\bot$ or $\|$ on the processes indicate the direction of the
magnetic field perpendicular or parallel to the momentum of the
initial particles.  $y$
is the sine of the scattering angle, $y^2=1-z^2$.  }
\end{center}
\end{table}

\begin{center}
\section*{\large\bf Figures}
\end{center}

\end{document}